\documentclass[a4paper,12pt]{article}
\usepackage{hyperref}
\usepackage{amsmath}
\usepackage{amssymb}
\usepackage{amsthm}
\usepackage{graphicx}

\begin{document}

\title{Translation Invariant Bipolaron Theory of Superconductivity and Spectroscopic Experiments}
\author{Victor D. Lakhno\thanks{lak@impb.ru} \\
Keldysh Institute of Applied Mathematics, \\
Russian Academy of Sciences, \\ 125047, Moscow, Russia
}
\date{}
\maketitle

\begin{abstract}
The explanation of the nature of superconducting gap in high temperature superconductors (HTSC) is a fundamental task which solution can lead to the understanding of superconducting mechanism. However, it has not been fully solved yet. From the mid of the twentieth century when Bardeen, Cooper and Schrieffer constructed their theory it has been believed that a superconducting gap is a collective phenomenon of electron excitations. In this work it is demonstrated that according to translation-invariant bipolaron theory of HTSC the different types of experiments measure for the gap different values. Thus tunneling experiments determine the bipolaron energy for a superconducting gap. On the other hand, the angle - resolved photoemission spectroscopy method measures the phonon frequency for which the electron-phonon interaction is maximum. Such effects as kinks in spectral measurements of gap, its angular dependence, existence of pseudogap and others have got natural explanations.
\end{abstract}

\vspace{10pt}
\vspace{2pc}
\section{\label{1-intro}Introduction }

In papers \cite{1,2} we constructed a translation invariant (TI) bipolaron theory of superconductivity (SC)
on the basis of a Hamiltonian of Pekar-Fr\"ohlich electron-phonon interaction (EPI) when EPI,
as distinct from the case of the Bardeen-Cooper-Schrieffer theory \cite{3}, cannot be considered
to be weak. The role of Cooper pairs in this theory belongs to TI bipolarons whose correlation
length ($\approx 1$nm) is much less than that of Cooper pairs ($\approx 10^3$ nm). According to \cite{1,2},
in high-temperature superconducting (HTSC) materials, TI bipolarons are formed near the
Fermi surface and represent a charged Bose gas capable of experiencing Bose-Einstein
condensation (BEC) at high critical temperature which determines the temperature of SC transition.

As distinct from Cooper pairs, TI bipolarons have their own excitation spectrum which determines
the thermodynamic properties of Bose gas of TI bipolarons. This spectrum has a gap which is equal
to the frequency of an optical phonon $\omega_0$  in an isotropic medium. In the case of weak EPI
(BCS limit), the frequency of an optical phonon is $\omega_0>>|E_B|$,
in the case of intermediate coupling  it is  $\omega_0\cong|E_B|$,
and in the case of strong coupling it is  $\omega_0<<|E_B|$,
where  $|E_B|$ is the energy of a TI bipolaron. According to \cite{2},
the number of TI bipolarons $N_B$ is: $N_B\approx N|E_B|/2E_F$,
where $N$ is the total number of electrons, $E_F$ is the Fermi energy, i.e. $N_B<<N$.
This means, in particular, that when calculating the London penetration depth
in the TI bipolaron theory in order to find the concentration of superconducting
current carriers at low temperatures, one should use $N_B$, rather than
the total number of current carriers $N$ as one does in the BCS.
For high-temperature superconductors, this fact was proved experimentally in \cite{4},
where London penetration depth was measured in overdoped SC.

The theory developed in \cite{1,2} fits in well with the thermodynamic properties of HTSC
materials and their magnetic characteristics \cite{5}. However, these facts alone do not enable
us to conclude that the TI bipolaron theory of SC does not contradict other experimental facts.

At present there are a lot of methods which enable one to study the properties of paired
states and the consequences arising from them. The aim of this work is to analyze to what
extent the data of contemporary spectroscopy methods, including  scanning tunneling
microscopy (STM), its attendant methods of quasiparticle interference, angle-resolved
photoemission spectroscopy (ARPES) and Raman (combination) scattering are compatible
with the ideas of TI bipolaron mechanism of HTSC.

The paper is arranged as follows. In Section~\ref{2-tunnel} we discuss the peculiarities of tunnel
experiments resulting from the properties of TI bipolarons. It is shown that
the presence of kinks in the spectral characteristics and conductance measurements
is a natural consequence of the TI bipolaron theory.

In Section~\ref{3-arpes} we discuss the data of ARPES experiments. It is shown that these data correspond
to the fact that the SC gap has a phonon nature as it follows from the TI bipolaron theory of SC.

In Section~\ref{4-raman} the data of combination scattering are compared with the conclusions arising from the TI bipolaron theory.

In Section~\ref{5-concl} we discuss the results obtained.

\section{\label{2-tunnel}Tunnel characteristics}

\begin{figure}[t]
\includegraphics{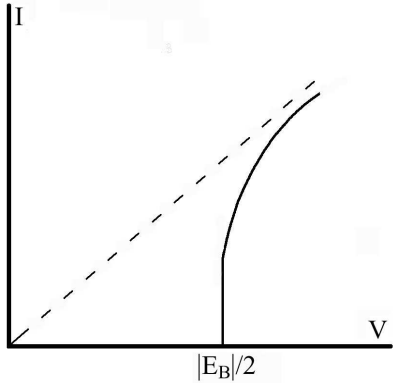}
\caption{\label{fig1}Current-voltage characteristic of one-particle current.}
\end{figure}

In the case of the TI bipolaron theory of SC, tunnel characteristics have some peculiarities.
As usual, in considering tunnel phenomena, for example, in considering a transition from
a superconductor to an ordinary metal
through a tunnel contact we will reckon the energy
from the ground state of a SC. In the TI bipolaron theory, the ground state is the bipolaron
one whose energy is lower than the Fermi level of this SC in normal state by the value of the bipolaron energy $|E_B|$.
Hence, upon a tunnel contact of  a SC with an ordinary metal, the Fermi level of an ordinary metal
will coincide with the ground state energy of a SC. Therefore the one-particle current for such
a contact will have the usual form (figure~\ref{fig1}).

\begin{figure}[t]
\includegraphics{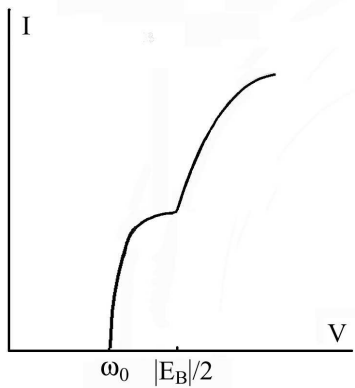}
\caption{\label{fig2}Current-voltage characteristic of total current.}
\end{figure}

The peculiarity arises in considering a two-particle current. It is concerned with the fact
that the spectrum of excited states of a TI bipolaron is separated from the ground state
by the value of the phonon frequency $\omega_0$. For this reason the current-voltage characteristic
of a two-particle current will have the form shown in figure~\ref{fig1} with $|E_B/2|$ replaced by $\omega_0$.
As a result the total current-voltage characteristic will have the form shown in figure~\ref{fig2}.
 	
The I-V curve represents the case of $\omega_0<|E_B|$. In the opposite case the quantities $\omega_0$
  and $|E_B|/2$ should be reversed. The region of the I-V curve determined by the values of $\omega_0<V<|E_B|/2$
	corresponds to the kink which is absent in the BCS theory.

\begin{figure}[t]
\includegraphics{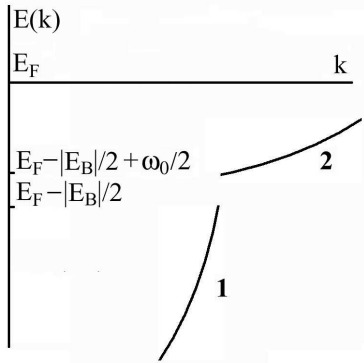}
\caption{\label{fig3}The kink corresponds to the gap which arises in passing from normal branch 1
to TI bipolaron branch 2 for energy equal to $E_F-|E_B|/2$.}
\end{figure}

From the spectral viewpoint, a kink corresponds to a transition from a one-particle
spectrum of electrons with energy lying below $E_F$ by the value of $|E_B|/2$  to a two-particle
TI bipolaron spectrum of excited states which in the one-particle scheme lies in
the region of \\ ($E_F-|E_B/2|+\omega_0/2$, $E_F$), as it is shown in figure~\ref{fig3}.

The dependence of $E(k)$  shown in figure~\ref{fig3} corresponds to ARPES observations of kinks
in a lot of HTSC materials (see, for example, review \cite{6}). For example, according to \cite{6},
in the well-studied cuprate Bi2212, the kink energy ($|E_B|/2$) is 70 meV.

\begin{figure}[t]
\includegraphics{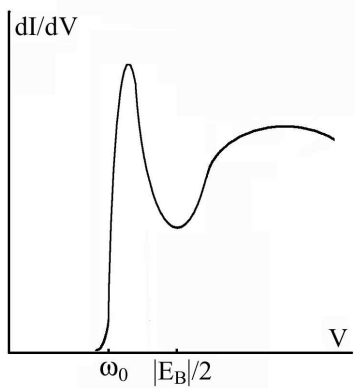}
\caption{\label{fig4}Dependence of $dI/dV$ conductance on $V$ corresponding to the curent-voltage characteristic  shown in figure~\ref{fig2}.}
\end{figure}

Figure~\ref{fig4} demonstrates a typical for HTSC dependence of $dI/dV$ on $V$ shown in figure~\ref{fig2}.
Here a kink corresponds to the dip to the right of  the high peak.

Notice that, in view of the fact that TI bipolarons exist even at $T>T_c$, the $dI/dV$
curve at $T>T_c$  will qualitatively retain the form shown in figure~\ref{fig4}.  Hence, the quantity $|E_B|/2$
will play the role of a pseudogap in one-particle transitions, while $|E_B|$ - the role
of a pseudogap in two-particle transitions. This conclusion is fully compliant
with numerous tunnel experiments in HTSC \cite{6,7,8}.
 	
\section{\label{3-arpes}Angle-resolved photoemission spectroscopy (ARPES)}

Apart from STM, a direct method providing information on the properties of a
superconducting gap is the angle-resolved photoemission spectroscopy \cite{9}.
When supported by the results of STM and evidence of quasiparticle interference
 this method provides the fullest data on the properties of a SC gap. Recently
a method of double photoelectron spectroscopy has been developed which is a generalization
 of ARPES to the case of two particles where two electrons with certain moments $\vec{k}_1$  and
$\vec{k}_2$  and relevant energies $E_1$  and  $E_2$  are photo-emitted \cite{10}.
Despite the abundance of data obtained by ARPES the nature of HTSC gap is still unclear.
In large part this is due to the fact that so fat a unified theory of HTSC has not been created.
On the understanding that a SC mechanism is caused by Cooper pairing, in the case
of strong Froehlich EPI this leads to the TI bipolaron theory of HTSC \cite{1,2}.
As distinct from bipolarons with broken symmetry, TI bipolarons are delocalized
in space and do not have a polarization potential well (polarization charge is equal to zero).
According to \cite{1,2}, a TI bipolaron has a gap in the spectrum which has a phonon nature.
In the TI bipolaron theory, bipolarons are formed near the Fermi surface in the form of
a charged Bose gas (immersed into electron gas) which is condensed at the level lying
lower than the Fermi level by the value equal to the ground state bipolaron energy which
leads to the SC state. The excitation spectrum of such a gas has a gap equal to the phonon frequency.
In this section we will show that the photoemission spectrum obtained in ARPES contains
exactly the kind of a gap, while the gap $|E_B|/2$  determined from the two-particle current
which was considered in the previous section has nothing to do with the gap measurements by ARPES.

With this end in view we will proceed from the general expression for the intensity of light
absorption $I(\vec{k},\omega)$  measured in ARPES such that:
\begin{eqnarray}
\label {eq.1}
I(\vec{k},\omega)=A(\vec{k},\omega)F(\omega)M({k},\omega).
\end{eqnarray}
In the case of intensity of light absorption by TI bipolarons measured by ARPES,
the meaning of the quantities involved in \eqref{eq.1} are different from that in
the case of one-electron photoemission.

In the case of Bose condensate under consideration, $\vec{k}$  has the meaning of the boson momentum,
$\omega$   is the boson energy,  $A(\vec{k},\omega)$ is a one-boson spectral function,
$F(\omega)$   is a function of Bose-Einstein distribution,
$M(\vec{k},\omega)$   is a matrix element describing transitions from
the boson initial state to its final state.

In our case the role of a charged boson taking part in the light absorption belongs to a bipolaron whose energy spectrum is determined by the expression \cite{1,2}:
\begin{eqnarray}
\label {eq.2}
 \epsilon_k=E_B\Delta_{k,0}+\left(E_B+\omega_0(\vec{k})+\frac{k^2}{2M}\right)\left(1-\Delta_{k,0}\right),
\end{eqnarray}
where  $\Delta_{k,0}=1$, if  $k=0$,  $\Delta_{k,0}=0$, if  $k\neq0$,
and whose distribution function equals \\ $F(\omega)=\left[\exp(\omega -\mu)-1\right]^{-1}$.
For $\vec{k}=0$, a TI bipolaron occurs in the ground state, while for $\vec{k}\neq0$ - in the excited
state with energy  $E_B+\omega_0(\vec{k})+k^2/2M$, where $\omega_0(\vec{k})$  is the phonon
frequency depending on the wave vector, $M=2m$, $m$ is the electron effective mass.

\begin{figure}[bt]
\includegraphics{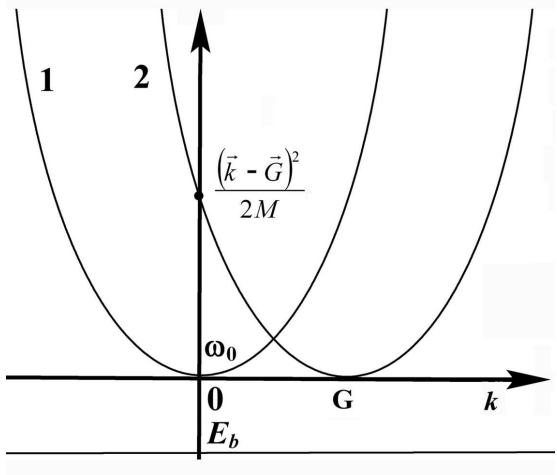}
\caption{\label{fig5}Schematic representation of a bipolaron transition into an excited state as a result of absorption of a quantum of light.}
\end{figure}

For further analysis it is important to note that the energy of bipolaron excited
states reckoned from $E_B$ in equation \eqref{eq.2} can be interpreted as a phonon energy $\omega_0(\vec{k})$
and a kinetic energy of two electrons coupled with this phonon. In the scheme of
expanded bands the latter has the form: $(\vec{k}+\vec{G})^2/2M$, where $\vec{G}$
is the reciprocal vector of the lattice (figure~\ref{fig5}). In ARPES technique measurements are taken
of the spectrum of initial states which is the spectrum  of low-lying excitations of a
TI bipolaron. In this connection we can neglect the contribution into this spectrum of
one- and two-particle excitations of an electron (polaron) gas into which the
bipolarons are immersed, since the density of the states of TI bipolarons near their
ground state is much greater than the density of the states of the electron spectrum.
Hence, we deliberately exclude consideration of such phenomena as de Haas--van
Alphen and Shubnikov--de Haas oscillations \cite{11,12,13}. Since the kinetic energy
corresponding to the reciprocal lattice vector (or the whole number of the reciprocal
lattice vectors) is very high, out of the whole spectrum of a bipolaron determined by
\eqref{eq.2} we should take account only of the levels $E_B$ with $k=0$ and $E_B+\omega_0(\vec{k})$
 with  $k\neq0$ as a spectrum of the initial states. In other words,
with the use of the spectral function $A(\omega ,\vec{k})=-(1/\pi)$Im $G(\omega ,\vec{k})$,
where $G(\omega ,\vec{k})=(\omega -\epsilon_k-i\epsilon)^{-1}$ is the bipolaron Green function,
the expression for intensity \eqref{eq.1} can be presented in the form:
\begin{eqnarray}
\label {eq.3}
I(\vec{k},\omega)\propto\frac{1}{(\omega -E_B)^2+\epsilon^2_1}\cdot
\frac{1}{(\omega-E_B-\omega_0(\vec{k}))^2+\epsilon^2_2},
\end{eqnarray}
which is fitting of the distribution function $F$ with $\mu=E_B$ and Green function $G$ with
the use of Lorentzians, where $\epsilon_1$ and  $\epsilon_2$ determine the width of
the Bose distribution and bipolaron levels, respectively (matrix element $M(\vec{k},\omega)$,
involved in \eqref{eq.1}, has a smooth dependence on the energy and the wave vector,
therefore this dependence can be neglected).

Hence, as a result of absorption of light by a pair of electrons (which initially occur in the bipolaron state)
ARPES method measures the kinetic energy of electrons with momenta $k_e$, which escape from
a sample into vacuum as a result of absorption of a photon with energy $\hbar\nu$.
The energy conservation law in this case takes the form:
\begin{eqnarray}
\label{eq.4}
\hbar\nu=\omega_0(\vec{k})+\frac{(\vec{k}+\vec{G})^2}{2M}=\zeta +\frac{k^2_e}{m_0}+\omega_0(\vec{k}),
\end{eqnarray}
$$\zeta=2\Phi_0+\left|E_B\right|,$$

\begin{figure}[t]
\includegraphics{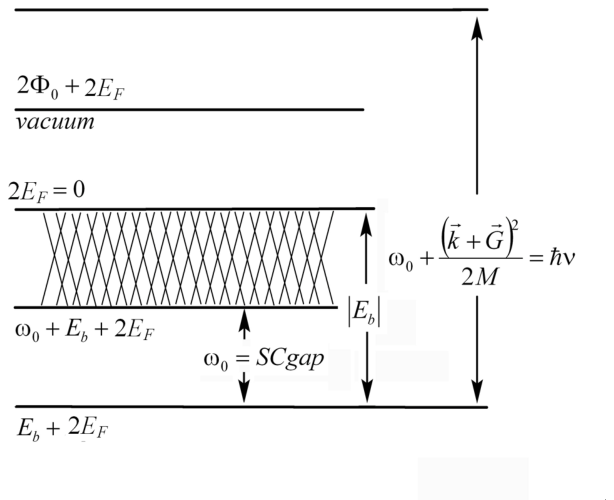}
\caption{\label{fig6}Scheme of energy levels in measuring the spectrum by ARPES. The area of the continuous spectrum lying below the Fermi level is crosshatched.}
\end{figure}

which is illustrated by figure~\ref{fig6}, where $\Phi_0$ is the work of electrons' escape from a sample,
$m_0$  is the mass of a free electron in vacuum. According to figure~\ref{fig6}, when a
bipolaron is formed near the Fermi energy $E_F$ the energy of two electrons becomes equal to $2E_F+E_B$.

In this case the electrons pass on from the state with $p_F$ (where $p_F$ is the Fermi momentum)
to a certain state with momentum $p$ below the Fermi surface (since $E_B<0$).
In ARPES method measurements are taken of the spectrum of initial states reckoned from
the energy $2\cdot p^2/2m$ corresponding to the energy of two electrons with momentum $p$.
As a result, in ARPES the energy $\omega_i=2E_F+E_B-p^2/m$  is measured.

Hence, if a bipolaron with energy $\omega=\omega_i=E_B+2\nu(\vec{p}_F-\vec{p})$
lying in the domain of existence of a bipolaron gas  $(2E_F+E_B,2E_F)$, where $\vec{\nu}_F$
is the velocity of a Fermi electron, absorbs a photon with energy $\hbar\nu$, then a phonon
arising as a result of a decay of a bipolaron is recorded in ARPES as a gap $\omega_0(\vec{k})$,
and two electrons with kinetic energy $k^2_e/m$  determined by \eqref{eq.4} are emitted from a sample.

In this scenario in each act of light absorption, two electrons with similar
momenta are emitted from a sample. This phenomenon can be detected by ARPES if
we place the electron detector just on the sample surface, since the kinetic energy of
the dispersion of two electrons in vacuum (not compensated by the attracting
potential in the bipolaron state) is several electron volts.

\begin{figure}[bt]
\includegraphics{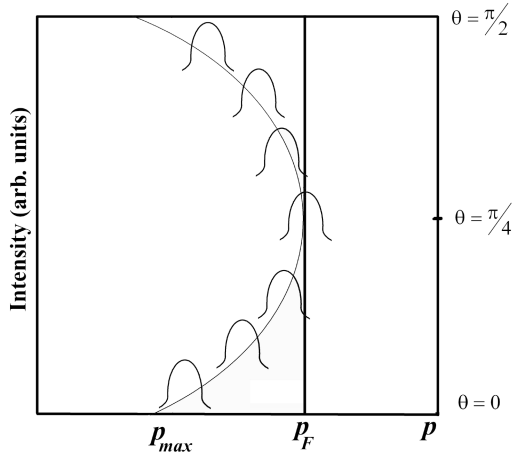}
\caption{\label{fig7} Schematic representation of the angular dependence of the absorption intensity determined by \eqref{eq.3} for $\omega =\omega_i$.}
\end{figure}

Hence, ARPES measures the phonon frequency $\omega_0(k)$, which is put in correspondence
with a SC gap and, therefore, in cuprate HTSC with $d_{x^2-y^2}$ symmetry its angular
 dependence is determined by the expression  $\omega_0(\vec{k})=\Delta_0|\cos k_xa-\cos k_ya|$.
This angular dependence $\omega_0(k)$  leads to the angular dependence of the intensity
$I(\omega_i,\vec{p})\propto A(\omega_i,\vec{p})$ given by equation \eqref{eq.3} (figure~\ref{fig7}) which is
usually observed in ARPES experiments \cite{9,14,15}. The form of the dependence $I(\omega_i,\vec{p})$
implies that there also exists a dependence of absorption peaks on $\vec{p}$  symmetric with
respect to the Fermi level. This dependence is not shown in figure~\ref{fig7} since in view of low
occupation density of states with $p>p_F$ their absorption intensity will be very small \cite{16}.

Experimental verification of the emission effect of TI bipolarons as a whole is
important to understand the pairing mechanism. Thus, according to \cite{15}, only one
electron should escape from a sample with dispersion of initial states determined,
by the formula: \\ $\epsilon^{Bog}_p=\sqrt{(p^2/2m-E_F)^2+\Delta^2(p)}$
(where $\epsilon^{Bog}_p$ is a spectrum of a Bogolyubov quasiparticle), different from spectrum \eqref{eq.2}.

The use of spectra $\epsilon^{Bog}_p$  and \eqref{eq.2} to describe the angular dependence of the intensity leads to a qualitative agreement with ARPES data with currently attainable resolution. Experiments with higher resolution should provide an answer to the question of whether the SC condensate in cuprates has fermion or TI bipolaron character.

\begin{figure}[bt]
\includegraphics[height=8cm]{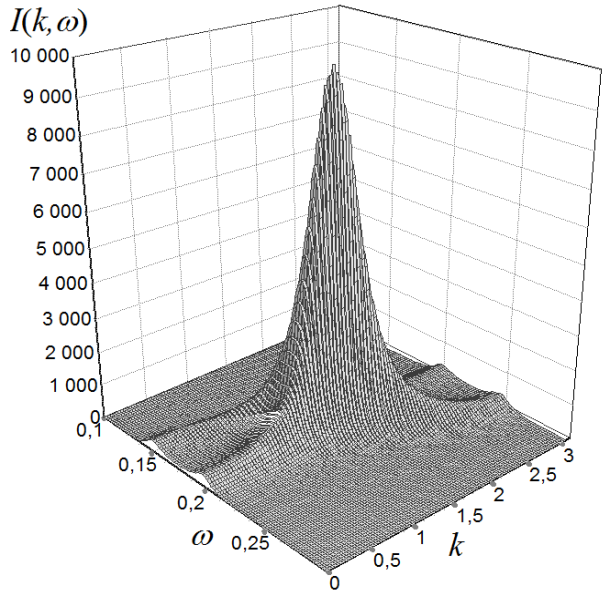}
\caption{\label{fig8} Dependence of the absorption intensity $I(k,\omega)$ (arbitrary units) on $k$  and
$\omega$  (eV) determined by \eqref{eq.3} for the parameters: $|E_B|=0,2$ eV, $\Delta_0=0,05$ eV,
$\epsilon_1=\epsilon_2=0,01$  eV and the wave vector $k$  in antinodal direction.
The lattice constant is assumed to be equal to one.}
\end{figure}

The spectrum of $\omega_0(k)$  suggests that in the nodal direction of cuprate superconductors the EPI constant becomes infinite. Hence, in this case for bipolarons, the strong coupling regime takes place. Figure~\ref{fig8} illustrates a typical dependence of the absorption intensity $I(\omega_i,\vec{p})$ observed in ARPES experiments \cite{14}.

The dependence observed in figure~\ref{fig8} follows from the intensity expression \eqref{eq.1}
with the spectral function which corresponds to the spectrum of a TI bipolaron \eqref{eq.2} and which
cannot be obtained from the spectral function \eqref{eq.3} from paper \cite{16} where Bogolyubov spectrum
$\epsilon^{Bog}_p$ is used for the spectrum and Fermi distribution function is used instead
of Bose distribution. This result can be considered as an argument in favour of a TI bipolaron mechanism of SC.

The above-considered peculiarities in the ARPES absorption spectrum will also manifest
themselves in tunnel experiments in the form of a thin structure (kinks) on the current-voltage
characteristics measured (Sec.~\ref{2-tunnel}). To this end, as distinct from traditional ARPES measurements
with high-energy photon sources ( $\hbar\nu=20-100$ eV), for these peculiarities to be observed one should
use sources with low-energy photons ($\hbar\nu=6-7$ eV) and higher than ordinary
resolution with respect to the momenta \cite{17,18,19,20}.

In paper \cite{21} a possibility to observe emission of Cooper pairs by ARPES was considered theoretically
for ordinary SC. In particular, the authors showed the presence of a peak in the emission current
of Cooper pairs which corresponds to zero coupling energy of occupied two-electron states.
The peak considered in \cite{21} corresponds to a transition with the energy $\hbar\nu$  determined by
\eqref{eq.4} and the coupling energy $\approx 1$ meV which is on the verge of the ARPES measurement accuracy.
In the case of high-temperature superconductors the coupling energy can be tenfold higher
which makes verification of the effects under consideration in them more realistic.
The main distinction of the results obtained in them from those obtained in \cite{21} is
the presence of an angular dependence of the absorption peak (figures~\ref{fig6},~\ref{fig7}) typical for HTSC materials.

Let us discuss briefly the temperature dependence of the intensity  $I(\omega_i,\vec{k})$.
According to equation \eqref{eq.1}, it is determined by the temperature dependence of $F(\omega)$.

For $T<T_c$, where $T_c$ is the temperature of a SC transition, $F\cong N_0(T)$ for $\omega=E_B$,
where $N_0(T)$ is the number of bosons
(bipolarons) in a condensate which determines the temperature dependence of the absorption
intensity. The value of  $N_0(T)$ decreases as $T$ grows and, generally speaking, vanishes at
the temperature of a SC transition making the absorption intensity equal to zero.
Actually, however, this is not the case, since only the Bose condensate side turns to zero.
According to TI bipolaron theory of SC, at   $T>T_c$ bipolarons exist in the absence of
a condensate too. In this case the population density of the ground state of such
bipolarons will decrease as the temperature grows vanishing at temperature $T^*$ corresponding
to a transition from the pseudogap to normal state.

This conclusion is confirmed by ARPES experiments in SC and pseudogap phases \cite{22} which
demonstrated that the angular dependence of a SC gap of \textit{d}-type is similar to the angular
dependence of the population density in the pseudogap phase. At the same time there
are considerable differences between the ARPES experimental data in SC and pseudogap phases.
In the SC phase the absorption intensity peak occurs below the Fermi level which corresponds
to the sharp  spectral peak of the population density of Bose condensate determined
by equation \eqref{eq.3} while in the pseudogap phase this peak will be absent in view of
the absence of a condensate in it \cite{23}. Under these conditions, in view of an increase
in the population density of excited bipolaron states with rising temperature,
the intensity of the absorption peak in ARPES experiments will decrease with a rise
in temperature and reach minimum in antinodal direction and maximum in the nodal one.

\section{\label{4-raman}Raman scattering}

Notwithstanding the fact that Raman scattering does not give an angular resolution \cite{24},
the data obtained by this method also point to the phonon nature of a gap in HTSC.
As was shown in \cite{1,2}, the spectrum determined by \eqref{eq.2} can be interpreted as a spectrum
of renormalized phonons. Scattering of light with frequency $\nu$  by such phonons will give
rise to satellite frequencies $\nu^B_+=\nu+|\epsilon^B_k|$  and  $\nu^B_-=\nu-|\epsilon^B_k|$
in the scattered light, where  $\epsilon^B_k$ is determined by \eqref{eq.2}.
In the case of wide conductivity bands, i.e. when inequality $G^2/M>>\max\omega_0(k)$  holds,
the splitted lines $\nu^B_\pm$  overlap and form a region with a maximum displaced to the Stokes branch $\nu_-$.
Since in the model under consideration a bipolaron gas is placed in a polaron gas where the number
of bipolarons is far less than the number of polarons, the intensity of bipolaron satellites
will be much weaker than that of TI polaron satellites: $\nu^P_+=\nu+|\epsilon^P_k|$  and
$\nu^P_-=\nu-|\epsilon^P_k|$, $\epsilon^P_k=E_P\Delta_{k,0}+\left(\omega_0+k^2/2m\right)\cdot\left(1-\Delta_{k,0}\right)$,
$E_P$  is the energy of a TI polaron.
As in the case of an ordinary combination scattering, the intensity of scattering on polarons and
bipolarons will be much weaker than the intensity of Rayleigh scattering corresponding to frequency $\nu$.

In experiments on Raman scattering \cite{25}, indeed, at $T<T_c$ a wide peak appears which, according to
our interpretation, corresponds to widen frequencies $\nu^{B,P}_{\pm}$. In full compliance with the experiment,
the position of this peak is independent of temperature. In the Raman scattering theory
based on the BCS, on the contrary, the position of the peak should correspond to the width
of the SC gap and at $T=T_C$  the frequency corresponding to this width should vanish.

The results of Raman scattering also confirm that TI bipolarons do not decay at  $T=T_C$, and still
exist in the pseudogap phase. Measurement of the temperature dependence of the Raman scattering
intensity is based on subtraction of the absorption intensity in normal and superconducting phases.
The remainder obtained, according to our approach, is totally determined by scattering on
the Bose condensate and, in full conformity with the experiment, is temperature-dependent, vanishing at  $T=T_C$.

\section{\label{5-concl}Conclusive remarks}

In this paper we practically did not use any specificity of the mechanism of pairing of electrons and holes.
For example, both in the Hubbard model and in the t-J model in describing HTSC copper oxides
the same holes take part both in the formation of antiferromagnetic fluctuations and in pairing
caused by exchange by these fluctuations. If an interaction of holes with magnetic fluctuations
leads to a formation of TI magnetopolarons having the spectrum $\omega_0(k)$, then this spectrum is also
the spectrum of magnons renormalized by their interaction with holes (bound magnons).
Exchange by such magnons can lead to the formation of hole pairs, i.e. formation of magnet
TI bipolarons whose spectrum, according to the TI bipolaron theory, will be just the spectrum
of  $\omega_0(k)$, i.e. the spectrum possessing $d$-symmetry. For this reason the assertion that a RVB superconductor
is merely a limit case of a BCS SC with strong interaction becomes justified \cite{26}.

Obviously, $d$-symmetry is specificity of cuprate HTSC and is not an indispensable condition for
HTSC existence. For example, such a substance as sulphur sulphide (H$_3$S) demonstrates a record
transition temperature: $T_C=203$ K (under high pressure \cite{27}), it does not have a magnetic order
but in it there is a strong EPI. Still greater value of $T_c$ under high pressure has recently
been obtained in $LaH_{10}$  with $T_C=260$K \cite{28}, it is also characterized by a strong EPI and a lack of magnetic order.

In conclusion it should be noted that the problem of pairing mechanism is still unclear. If this is interaction of current carriers with magnetic fluctuations, then within our approach, the particles which will twin electrons will be not phonons but magnons. In passing on from the pseudophase to the normal phase this glueing mode disappears which leads to a decay of a bipolaron into two individual polarons with an emission of a phonon (magnon).

In the pseudogap phase there may be a lot of different gaps caused by the presence of phonons, magnons, plasmons and other types of elementary excitations. In this case the SC gap will be determined by the type of elementary excitations having the maximum force of interaction with current carriers.


\begin{thebibliography}{99}

\bibitem{1} V. D. Lakhno, Adv. Cond. Mat. Phys. 2018, 1380986, (2018) \url{https://doi.org/10.1155/2018/1380986}.

\bibitem{2} V. D. Lakhno, Physica C: Superconductivity and its applications 561, 1-8, (2019) \url{https://doi.org/10.1016/j.physc.2018.10.009}.

\bibitem{3} J. Bardeen, L. N. Cooper and J. R. Schrieffer, Phys. Rev. 108, 1175 (1957) \url{https://link.aps.org/doi/10.1103/PhysRev.108.1175}.

\bibitem{4} I. Bo\v{z}ovi\'{c}, X. He, J. Wu, A. T. Bollinger, Nature 536, 309 (2016) \url{https://doi.org/10.1038/nature19061}.

\bibitem{5} V. D. Lakhno, Condens. Matter, 4, 43, (2019) \url{https://doi.org/10.3390/condmat4020043}.

\bibitem{6} D. R. Garcia and A. Lanzara, Adv. Cond. Mat. Phys. 2010, 807412 (2010) \url{https://doi.org/10.1155/2010/807412}.

\bibitem{7} F. Giubileo, D. Roditchev, W. Sacks, R. Lamy, D. X. Thanh, J. Klein, S. Miraglia, D. Fruchart, J. Marcus, and Ph. Monod, Phys. Rev. Lett. 87, 177008 (2001) \url{https://doi.org/10.1103/PhysRevLett.87.177008}.

\bibitem{8} F. Giubileo, D. Roditchev, W. Sacks, R. Lamy, J. Klein, EuroPhys. Lett. 58, 764 (2002) \url{https://doi.org/10.1209/epl/i2002-00415-5}.

\bibitem{9} A. Damascelli, Z. Hussain, and Z.-X. Shen, Rev. Mod. Phys. 75, 473 (2003) \url{https://link.aps.org/doi/10.1103/RevModPhys.75.473}.

\bibitem{10} M. Hattass, T. Jahnke, S. Sch\"ossler, A. Czasch, M. Sch\"offler, L. Ph. H. Schmidt, B. Ulrich, O. Jagutzki, F. O. Schumann, C. Winkler, J. Kirschner, R. D\"orner, and H. Schmidt-B\"ocking, Phys. Rev. B 77, 165432 (2008) \url{https://link.aps.org/doi/10.1103/PhysRevB.77.165432}.

\bibitem{11} B. Vignolle, A. Carrington, R. A. Cooper, M. M. J. French, A. P. Mackenzie, C. Jaudet, D. Vignolles, C. Proust, and N. E. Hussey,
Nature, 455, 952 (2008) \url{https://doi.org/10.1038/nature07323}.

\bibitem{12} E. A. Yelland, J. Singleton, C. H. Mielke, N. Harrison, F. F. Balakirev, B. Dabrowski, and J. R. Cooper, Phys. Rev. Lett. 100, 047003 (2008) \url{https://link.aps.org/doi/10.1103/PhysRevLett.100.047003}.

\bibitem{13} T. Helm, M. V. Kartsovnik, M. Bartkowiak, N. Bittner, M. Lambacher, A. Erb, J. Wosnitza, and R. Gross, Phys. Rev. Lett. 103, 157002 (2009) \url{https://link.aps.org/doi/10.1103/PhysRevLett.103.157002}.

\bibitem{14} S. V. Borisenko, A. A. Kordyuk, S. Legner, C. D\"urr, M. Knupfer, M. S. Golden, J. Fink, K. Nenkov, D. Eckert, G. Yang, S. Abell, H. Berger, L. Forr\'o, B. Liang, A. Maljuk, C. T. Lin, and B. Keimer, Phys. Rev. B 64, 094513 (2001) \url{https://link.aps.org/doi/10.1103/PhysRevB.64.094513}.

\bibitem{15} K. M. Shen, F. Ronning, D. H. Lu, W. S. Lee, N. J. C. Ingle, W. Meevasana, F. Baumberger, A. Damascelli, N. P. Armitage, L. L. Miller, Y. Kohsaka, M. Azuma, M. Takano, H. Takagi, and Z.-X. Shen, Phys. Rev. Lett. 93, 267002 (2004) \url{https://doi.org/10.1103/PhysRevLett.93.267002}.

\bibitem{16} H. Matsui, T. Sato, T. Takahashi, S.-C. Wang, H.-B. Yang, H. Ding, T. Fujii, T. Watanabe, and A. Matsuda, Phys. Rev. Lett. 90, 217002 (2003) \url{https://link.aps.org/doi/10.1103/PhysRevLett.90.217002}.

\bibitem{17} I. M. Vishik, W. S. Lee, F. Schmitt, B. Moritz, T. Sasagawa, S. Uchida, K. Fujita, S. Ishida, C. Zhang, T. P. Devereaux, and Z. X. Shen, Phys. Rev. Lett. 104, 207002 (2010) \url{https://link.aps.org/doi/10.1103/PhysRevLett.104.207002}.

\bibitem{18} N. C. Plumb, T. J. Reber, J. D. Koralek, Z. Sun, J. F. Douglas, Y. Aiura, K. Oka, H. Eisaki, and D. S. Dessau, Phys. Rev. Lett. 105, 046402 (2010) \url{https://link.aps.org/doi/10.1103/PhysRevLett.105.046402}.

\bibitem{19} H. Anzai, A. Ino, T. Kamo, T. Fujita, M. Arita, H. Namatame, M. Taniguchi, A. Fujimori, Z.-X. Shen, M. Ishikado, and S. Uchida, Phys. Rev. Lett. 105, 227002 (2010) \url{https://link.aps.org/doi/10.1103/PhysRevLett.105.227002}.

\bibitem{20} J. D. Rameau, H.-B. Yang, G. D. Gu, and P. D. Johnson, Phys. Rev. B 80, 184513 (2009) \url{https://link.aps.org/doi/10.1103/PhysRevB.80.184513}.

\bibitem{21} K. A. Kouzakov and J. Berakdar, Phys. Rev. Lett. 91, 257007 (2003) \url{https://link.aps.org/doi/10.1103/PhysRevLett.91.257007}.

\bibitem{22} G. Varelogiannis Phys. Rev. B 57, R732(R) (1998) \url{https://link.aps.org/doi/10.1103/PhysRevB.57.R732}.

\bibitem{23} M. R. Norman, D. Pines and C. Kallin, Advances in Physics, 54, 715 (2005) \url{https://doi.org/10.1080/00018730500459906}.

\bibitem{24} T. P. Devereaux and R. Hackl, Rev. Mod. Phys. 79, 175 (2007) \url{https://link.aps.org/doi/10.1103/RevModPhys.79.175}.

\bibitem{25} O. V. Misochko,  Phys. Usp. 46, 373 (2003) \url{https://doi.org/10.1070/PU2003v046n04ABEH001257}.

\bibitem{26} S. A. Kivelson and D. S. Rokhsar, Phys. Rev. B 41, 11693(R) (1990) \url{https://link.aps.org/doi/10.1103/PhysRevB.41.11693}.

\bibitem{27} A. P. Drozdov, M. I. Eremets, I. A. Troyan, V. Ksenofontov and S. I. Shylin, Nature 525, 73 (2015) \url{https://doi.org/10.1038/nature14964}.

\bibitem{28} M. Somayazulu, M. Ahart, A. K. Mishra, Z. M. Geballe, M. Baldini, Y. Meng, V. V. Struzhkin and R. J. Hemley, Phys. Rev. Lett. 122, 027001 (2019)  \url{https://link.aps.org/doi/10.1103/PhysRevLett.122.027001}.

\end{thebibliography}
\end{document}